\documentclass{article}

\usepackage{PRIMEarxiv}
\usepackage{longtable}
\usepackage[utf8]{inputenc} 
\usepackage[T1]{fontenc}    
\usepackage{hyperref}       
\usepackage{url}            
\usepackage{booktabs}       
\usepackage{amsfonts}       
\usepackage{nicefrac}       
\usepackage{microtype}      
\usepackage{lipsum}
\usepackage{fancyhdr}       
\usepackage{graphicx}       
\graphicspath{{media/}}     
\usepackage{txfonts}
\usepackage{xcolor}

\title{Bimodal distribution of the solar wind at 1AU
\thanks{\textit{\underline{Citation}}: 
\textbf{Larrodera, C. and C. Cid (2020a). “Bimodal distribution of the solar wind at 1 AU”. In:
Astronomy \& Astrophysics 635, A44. DOI: 10.1051/0004-6361/201937307.}} 
}

\author{
  C. Larrodera \\
  University of Alcalá \\
  Alcalá de Henares\\
  \texttt{carlos.larrodera@edu.uah.es} \\
   \And
  C. Cid \\
  University of Alcalá \\
  Alcalá de Henares\\
  \texttt{consuelo.cid@uah.es} \\
}

\begin{document}
\maketitle

\begin{abstract}
Here we aim to separate the two main contributions of slow and fast solar wind that appear at 1 AU.
The Bi-Gaussian function is proposed as the probability distribution function of the two main components of the solar wind. The positions of the peaks of every simple Gaussian curve are associated with the typical values of every contribution to solar wind. 
We used the entire data set from the Advanced Composition Explorer (ACE) mission in an analysis of the data set as a whole and as yearly series.
Solar cycle dependence is considered to provide more accurate results for the typical values of the different parameters.  
The distribution of the solar wind at 1 AU is clearly bimodal, not only for velocity, but also for proton density, temperature and magnetic field. New typical values for the main parameters of the slow and fast components of the solar wind at 1 AU are proposed.
\end{abstract}


\section{Introduction}

The observations of the solar wind made during the Mariner 2 flight to Venus in 1962 discovered streams of hot, high-velocity plasma recurring at intervals of 27 days \cite{Neugebauer1966}. Between streams the proton velocity dropped as low as 307 km s$^{-1}$ from an average value of approximately 500 km s$^{-1}$. These were just the first observations of the two main contributions to solar wind: the fast flow, from coronal holes, and the slow wind, with a less certain origin \cite{abbo2016}. Transient ejections with a wide range of velocities are also contributing to the solar wind observed at 1 AU. 

Typical features of slow and fast components of the solar wind at 1 AU have been summarised by several authors (see reviews by \cite{Bothmer2007} and \cite{Hansteen2010} and references cited therein).
Based on the statistical analysis of samples from different missions, all authors agree that fast wind is hotter and less dense than slow solar wind,  but some discrepancies appear when comparing the values of the basic parameters of both types of solar wind (see Table \ref{tab:solar_wind_contributions}). One of the differences is that magnetic field strength is the same for both contributions according to \cite{Hansteen2010} while some differences appear for \cite{Bothmer2007}. Although these differences are included in the uncertainty on the value provided by \cite{Hansteen2010}, it would be useful to know whether the magnetic field strength presents some differences between both components of solar wind.

\begin{table*}[h]
    \caption{Typical values for the parameters of the slow and fast components of the solar wind.}
    \label{tab:solar_wind_contributions}
    \centering
    \begin{tabular}{lccccl} 
   \hline\hline
     & $V_{p} (km/s)$ & $B (nT)$ & $n_{p} (cm^{-3})$  & $T_{p} (\times 10^{5} K)$ &  \\ 
     \hline
     Slow wind  & $\leq 450$  & 4  & 7 $-$ 10 & 0.4& \cite{Bothmer2007} \\    
     & 430 $\pm$ 100 & 6 $\pm$ 3  & $\simeq$ 10 & 0.4 $\pm$ 0.2&\cite{Hansteen2010} \\
   \hline
    Fast wind & 450 $-$ 800 & 5  & 3   & 2& \cite{Bothmer2007}  \\
     & 700 $-$ 900 & 6 $\pm$ 3 & $\simeq$ 3   & 2.4 $\pm$ 0.6& \cite{Hansteen2010}\\
    \hline
    \end{tabular}
\end{table*}

Comparing the interplanetary magnetic field along the solar cycle, \cite{Hirshberg1969} reported high tails in the distribution function of the  magnitude of the interplanetary magnetic field ($B$) during a period of rising solar activity (1966-1967). However, no tail was seen during solar minimum (1963-1964). The tails in the distribution functions of solar wind plasma and magnetic field were attributed to solar mass ejections and the compression of solar wind regions of different speeds \cite{Neugebauer1966, Hirshberg1969}.

\cite{Burlaga1979} were the first to suggest that the distribution of the logarithm of $B$ data could be well represented as normally distributed. The analysis of skewness and kurtosis of the $log B$ for years 1973 to 1990 allowed \cite{Feynman1994} to demonstrate that the distribution of logarithms of one-hour averaged $B$ is non-Gaussian. Nevertheless, the lognormal distribution has been  used extensively to describe not only $B$, but also solar wind speed, density and temperature \cite[e.g.][] {Burlaga1999, Burlaga2001, Veselovsky2010, Venzmer2018}.

The lognormal distribution was considered as proof of a lack of separation between fast and slow flows at 1 AU, describing the solar wind as a simple statistical structure resulting from the dynamical evolution and interaction of the flows at 1 AU \cite{Burlaga1999, Burlaga2001}. 
Moreover, \cite{Venzmer2018} noted that the lognormal distribution appears 
to be better at describing the shape of the IMF, the density and the temperature distributions than at describing the solar wind speed. Therefore these latter authors considered a bi-component lognormal for the velocity distribution with a better result.

Other approaches to the distribution function of magnetic field strength, or other solar wind plasma parameters, consist in using the kappa-like distributions, adding in some cases  artificial terms to make the symmetric distribution more skewed \cite{Burlaga2004}. Kappa-like distributions with fat tails can be obtained as a superposition of random uncorrelated, normally or lognormally distributed processes \cite{Voros2015} in the context of non- extensive (non-additive) statistical mechanics. Also a randomised Weibull probability distribution function (PDF) was proposed for the magnetic field intensity \cite{Consolini2009}.

The historical evolution described above regarding the analysis of the distribution function of the solar wind parameters shows that more complex mathematical functions are being implemented as time passes assuming that the mixing and dynamical interaction makes separating the different components of the solar wind at 1 AU impossible. Nevertheless, slow and fast solar wind intervals are clearly identified at 1 AU.  

In order to identify the solar sources of individual packets of solar wind, \cite{Stansby2019} used solar wind measurements closer to the Sun than 1 AU, where the mixing and dynamical interaction of different solar wind streams is reduced. After removing all the intervals identified as coronal mass ejections from the data, these latter authors found three different populations at 0.3 AU,   corresponding to wind that originated: (1) in the core of coronal holes, (2) in or near active regions or edges of coronal holes, and (3) in small transients. On the other hand, the first observations from the Parker Solar Probe at 36 to 54 solar radii show evidence of slow Alfvénic solar wind emerging from a small equatorial coronal hole \cite{bale2019}.  

In this paper we separate the two main components of solar wind at 1AU, fast and slow flows, from other contributions to the distribution function of the observed values of the main parameters of the solar wind measured by the Advanced Composition Explorer (ACE) spacecraft. The ACE data are described in Section 2, and the bi-Gaussian approach with the statistical results and the solar cycle dependence are presented in Section 3. Finally, Section 4 discusses the results and Section 5 summarises our conclusions.

\section{Data}

\begin{figure}
	\includegraphics[width=\columnwidth]{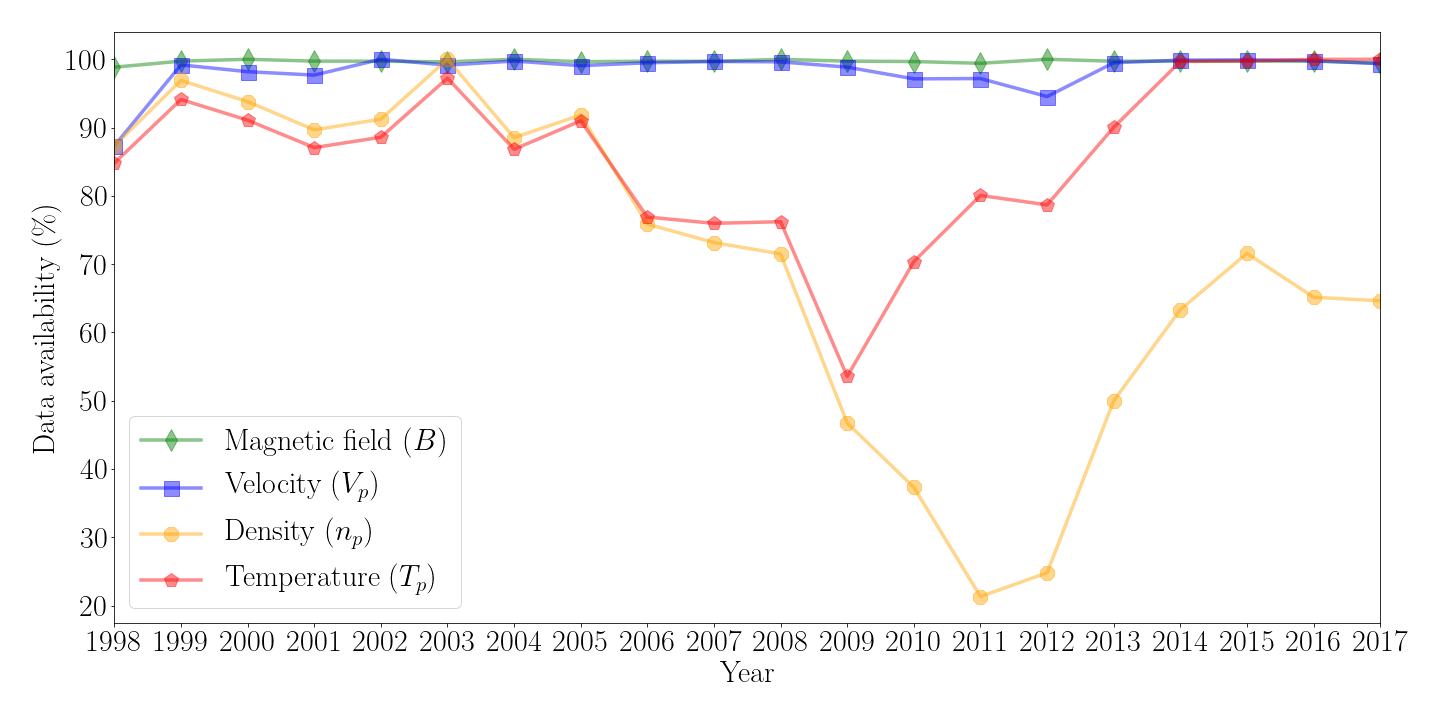}
    \caption{Data coverage of the main solar wind parameters measured by the ACE spacecraft from the time it is operational until the end of the year 2017.}
    \label{fig:number_data}
\end{figure}
This study uses data from the ACE spacecraft. The ACE mission orbits the L1 point, and has a prime view of the solar wind accelerated by the Sun. Data used here were measured by the Solar Wind Electron, Proton, and Alpha Monitor (SWEPAM; \cite{McComas1998}) and the Magnetic Field Experiment (MAG; \cite{Smith1998}) on board the ACE spacecraft. The level 2 data from SWEPAM and MAG were obtained from the ACE Science Center (\href{http://www.srl.caltech.edu/ACE/}{www.srl.caltech.edu/ACE/}). 
    
The ACE spacecraft provides continuous coverage of solar wind parameters since early 1998, although with some data gaps in SWEPAM data during periods of high solar activity. In this study we use the entire hourly data set from 23 January 1998 to December 2017 (about two solar cycles, namely 23 and 24). The data set includes the proton density, $n_{p}$, the radial component of the proton temperature, $T_{p}$ and the proton speed, $V_{p}$, from SWEPAM, and the magnetic field magnitude, $B$ from MAG.
    
The data coverage of different solar wind parameters is shown in Fig. \ref{fig:number_data}. We highlight the good coverage of the magnetic field magnitude and the proton speed many years past the original planned mission lifetime. Figure \ref{fig:number_data} also shows that, as informed by the SWEPAM instrument team, the proton density, and to a lesser extent the temperature, became increasingly sparse starting in 2010 as the detectors aged. Since 23 October 2012, an operational improvement has significantly increased the frequency of good-quality SWEPAM observations.

\section{A bi-Gaussian approach}

Considering the recent results from Parker Solar Probe \cite{bale2019}, both slow and fast contributions to solar wind emerge from coronal holes, small or large, respectively. Therefore, the simplest model for the solar wind consists in two flows from two coronal holes (one large and one small) with different speeds (one fast and one slow) emerging from the Sun. Nevertheless we should be aware that not all large (or small) coronal holes present the same size. Small changes in this size are expected to result in small changes in the speed of the emerging flow. Indeed, \cite{Garton_2018_Solar_Wind} show that the longitudinal width of coronal holes presents a strong correlation with the peak of the solar wind speed measured by ACE. Also the speed of a flow can be modified during the propagation until reaching 1 AU.

Here we use a bi-Gaussian approach as the probability distribution function of the two main components of solar wind at 1AU (slow and fast flows). This function provides the mathematical framework for the physical model described above.

The bi-Gaussian is defined as the addition of two Gaussian (or normal) distribution functions, one for the fast wind and another for the slow wind, as is given by 
    \begin{equation}
       bG(x)=h_{1}\cdot\mathrm{exp}{\left(\frac{-(x-p_{1})^{2}}{2w_{1}^{2}}\right)}+h_{2}\cdot\mathrm{exp}{\left(\frac{-(x-p_{2})^{2}}{2w_{2}^{2}}\right)} 
      \label{eq:bigauss}
    \end{equation}
where $x$ is $V_{p}$, $n_{p}$, $T_{p}$ or $B$ depending on the selected data set. Here, $h_{1}, p_{1}, w_{1}, h_{2}, p_{2}$ and $w_{2}$ are the six parameters obtained when fitting the bi-Gaussian distribution function to a data set. The subscripts $1$ and $2$ correspond to the normal distribution of each component of the solar wind. Be aware that $1$ and $2$ do not always correspond to slow and fast wind, respectively. Here $h$ is the height of the peak of each single Gaussian curve, $p$ is the position of the centre of the peak and $w$ is the Gaussian RMS width of each single curve. 

\subsection{Statistics with the whole data set}

\begin{figure}
	\includegraphics[width=\columnwidth]{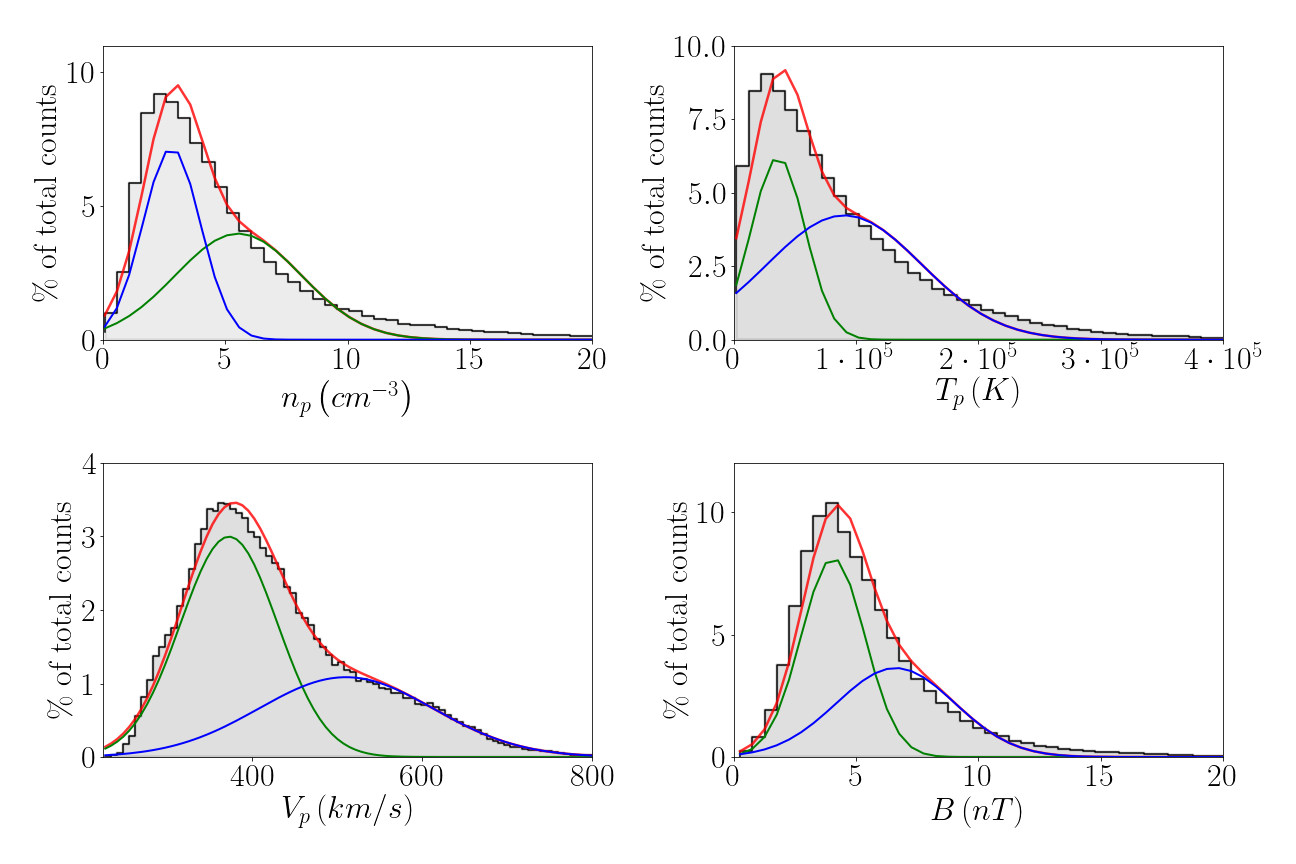}
    \caption{Empirical distribution functions of the main solar wind parameters, $n_{p}$, $T_{p}$, $V_{p}$, and $B$, for the whole data available from ACE and the fitting to a bi-Gaussian (red). Green and blue lines correspond to the single Gaussian curves.}
    \label{fig:biGaussian_all}
\end{figure}

Figure \ref{fig:biGaussian_all} shows the empirical distribution functions of the selected solar wind parameters, $V_{p}$, $n_{p}$, $T_{p}$ or $B$, for the whole data set available from ACE and the fitting to a bi-Gaussian. The two single Gaussian curves also appear as the blue and green lines. 

The Pearson chi-square (hereafter $chi^{2}$) has been used to quantify the goodness of fit, as the  size of the sample is large enough to accurately use this goodness of fit test. 
The  $chi^{2}$ is calculated using $chi^{2}=\sum^{k}_{i=1}{\left(O_{i}-E_{i}\right)^{2}/E_{i}}$, 
where $O_{i}$ is the number of values of a PDF falling into the i-th interval and $E_{i}$ is the expected number of values considering the theoretical function (Eq. \ref{eq:bigauss}) for the same interval  \cite{Suhov_2014_Probability}.
The $chi^{2}$ values obtained appear in the last column of Table \ref{tab:Parameters_all} , ranging from 0.02 for $V_{p}$ to 21.77 for $n_{p}$, indicating that the best (worse) fitting corresponds to the proton speed (density). In any case, the values obtained allow us to claim that the fitting of a Bi-Gaussian is appropriate for the PDFs of all the solar wind parameters analysed. This fact suggests that two solar wind populations can be distinguished, not only in the proton speed, but also in $n_{p}$, $T_{p}$ and $B$.

Most of the proton speed is spread around 370 km s$^{-1}$, but there is also a significant fraction of data spread around 510 km s$^{-1}$, corresponding to slow and fast contributions to the solar wind, as expected. Other values corresponding to the different solar wind parameters are in Table~\ref{tab:Parameters_all}.

\begin{table*}
\caption{Parameters obtained from the fitting of a Bi-Gaussian function to the proton density, the radial component of the proton temperature, the proton speed and the magnetic field strength from the ACE spacecraft since operations began up to 31 December 2017. The $chi^{2}$ of the fitting and the number of data points in every sample appear in the two last columns.}

\label{tab:Parameters_all}
\centering 
\begin{tabular}{lcccccccc} 
\hline\hline
 & $h_{1}$ & $p_{1}$ & $w_{1}$  & $h_{2}$ & $p_{2}$ & $w_{2}$ & $chi^{2}$ &  Data points \\ 
 \hline
$V_{p}$     & 3.00          & 373 $km/s$     & 57 $km/s$   & 1.09      & 510 $km/s$         & 103 $km/s$        & 0.02 & 172,156 \\
$T_{p}$     & 6.22          & 3.65 $\times 10^{4}K $     & 2.20 $\times 10^{4}K $ & 4.23      &  9.06$\times 10^{4} K$  & 6.31 $\times 10^{4} K$ & 3.59  & 148,379 \\
$n_{p}$ & 7.18          & 2.80 $cm^{-3}$             & 1.18 $cm^{-3}$  & 3.40      & 5.53 $cm^{-3}$           & 2.58 $cm^{-3}$           & 21.77 & 117,548 \\
$B$     & 8.12          & 4.06 $nT$                 & 1.31 $nT$   & 3.63      & 6.63 $nT$             & 2.42 $nT$            & 4.78  & 175,137\\		\hline
\end{tabular}
\end{table*}
 The tails, which are not completely reproduced by the bi-Gaussian, are expected to be related to solar transients and the compression of solar wind regions of different speeds \cite{Wimmer2006}. However, data collected during these time intervals are not statistically significant. Indeed, when computing how much the real distribution function departs from the Bi-Gaussian on the right side of the graphs (i.e. computing the amount of data which are left above the Bi-Gaussian on that side), we obtain values which range from  0.4\% for the solar wind velocity up to 9\% for the proton density.

To identify which one of the two Gaussian curves (1 or 2) corresponds to each component of the solar wind (slow or fast), we compare the position of the centre of the peak of every single Gaussian curve, $p$, with the values in Table \ref{tab:solar_wind_contributions}. The peak of the first (second) Gaussian curves of $V_{p}$, $B$ and $T_{p}$ fits well with the values for the slow (fast) contribution in Table \ref{tab:solar_wind_contributions}. On the contrary, the peak of the second (first) Gaussian curve of $n_{p}$ is the one which agrees with the slow (fast) contribution.

Thus, considering the position of the peak of every Gaussian as the value for a magnitude of a type of solar wind, with an uncertainty equal to the spread of the data around the peak, that is, equal to the RMS of the curve, we obtain the main parameters of each contribution to the solar wind at 1 AU (see Table \ref{tab:fit_values_results}).

\begin{table}
    \caption{Values for the main parameters of the slow and fast components of the solar wind at 1 AU from the bi-Gaussian fitting to the whole data set.}
    \label{tab:fit_values_results}
    \centering 
    \begin{tabular}{lcccc} 
    \hline\hline
     & $V_{p} (km/s)$ & $B (nT)$ & $n_{p} (cm^{-3})$  & $T_{p} (\times 10^{5} K)$  \\      \hline
    Slow wind   & $370\pm60$    & $4\pm1$   & $5\pm3$ & $0.4\pm0.2$\\    
    Fast wind   & $500\pm100$   & $6\pm2$   & $3\pm2$ & $0.9\pm0.6$\\    
    \hline
    \end{tabular}
\end{table}

\subsection{Yearly data sets and solar cycle dependence}

Here we decipher whether or not the values obtained for the two contributions of the main solar wind parameters (Table \ref{tab:fit_values_results}) change over time. Specifically we are interested in the dependence of the centre of the peak of the first and the second Gaussian curves ($p_{1}$ and $p_{2}$) on the solar cycle. For this purpose, we fitted the empirical yearly PDFs of $V_{p}$, $n_{p}$, $T_{p}$ and $B$ to a bi-Gaussian function.

\begin{table}
    \caption{Same as Table \ref{tab:fit_values_results} but in this case the values are obtained from the weighted average of the yearly peaks of the first and second Gaussian. Uncertainty corresponds to the weighted standard deviation}
    \label{tab:weighted values}
    \centering 
    \begin{tabular}{lcccc} 
    \hline\hline
     & $V_{p} (km/s)$ & $B (nT)$ & $n_{p} (cm^{-3})$  & $T_{p} (\times 10^{5} K)$  \\      \hline
    Slow wind   & $376\pm22$    & $4\pm1$   & $6\pm1$ & $0.5\pm0.3$\\    
    Fast wind   & $496\pm56$   & $7\pm1$   & $3\pm1$ & $1.0\pm0.4$\\    
    \hline
    \end{tabular}
\end{table}

\begin{figure}
	\includegraphics[width=\columnwidth]{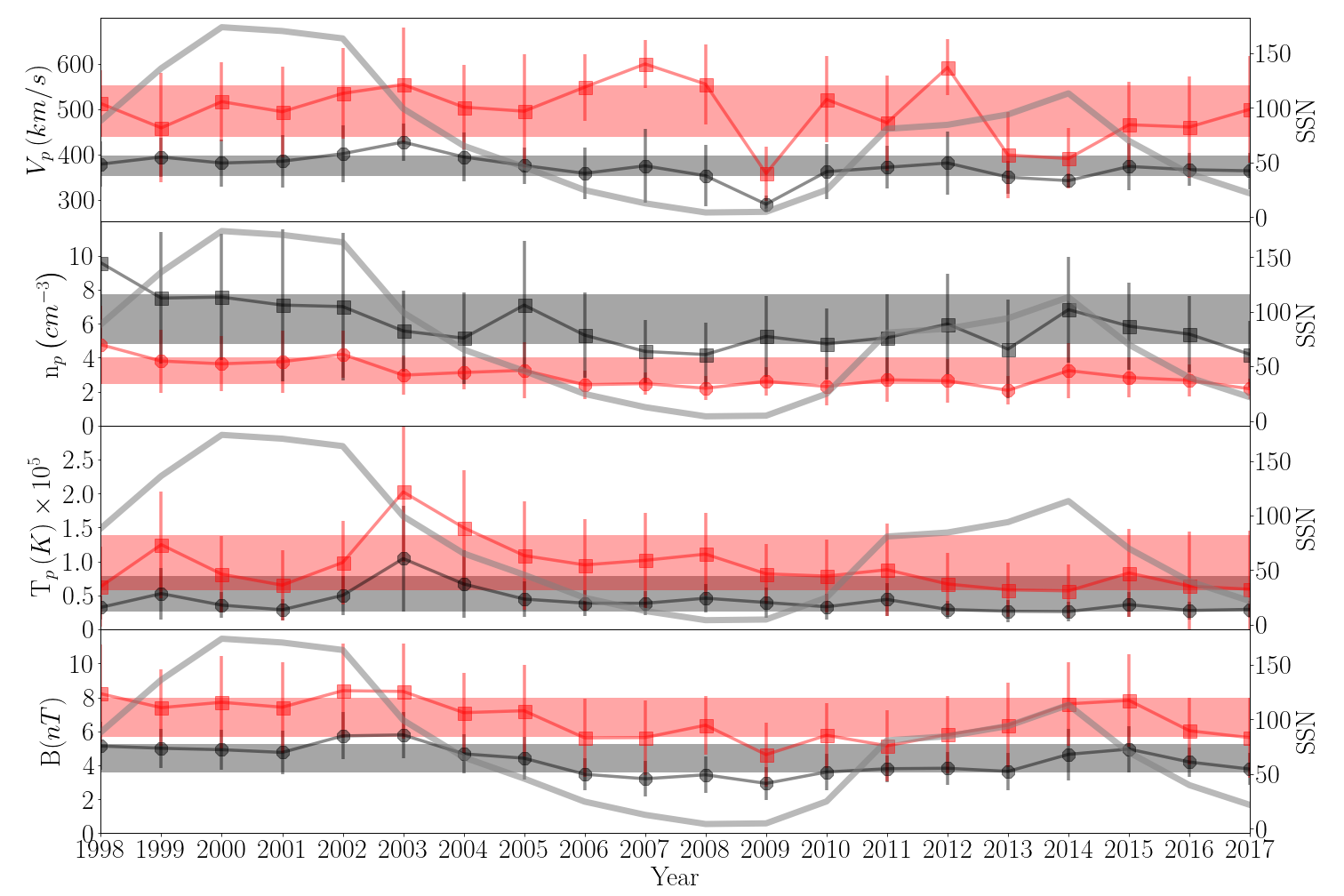}
    \caption{From top to bottom: Yearly position of the centre of the peak of every single Gaussian curve, $p$, for $V_{p}$, $n_{p}$, $T_{p}$ and $B$ PDFs. Black (red) points correspond to slow (fast) wind. Uncertainty has been estimated using the Gaussian RMS width of the corresponding single curve, $w$. The grey line in the four plots represents the sunspot number, with the corresponding $y$-axis on the right of every plot.}
    \label{fig:solarCycle_all}
\end{figure}

Figure \ref{fig:solarCycle_all} compares the evolution of $p_{1}$ and $p_{2}$ for the main solar wind parameters to the yearly sunspot number (SSN) from WDC-SILSO as a proxy of the solar cycle. Shadowed areas in the different panels are centred at the weighted average of the peak of corresponding Gaussian curve and extend to $\pm \sigma$ (see Table \ref{tab:weighted values}). Grey (red) colour corresponds to the slow (fast) contribution to solar wind. 

\begin{figure}
	\includegraphics[width=\columnwidth]{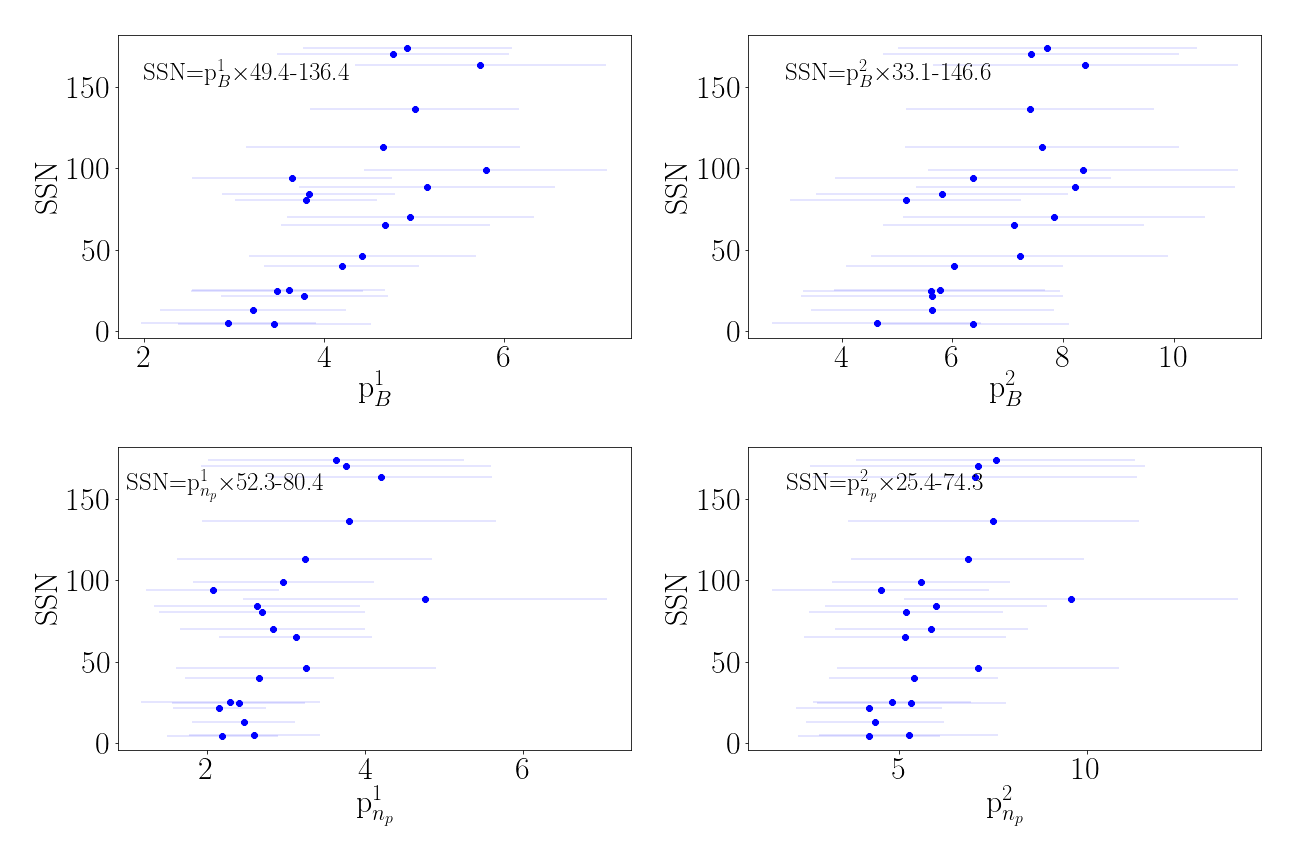}
    \caption{Scatter plots of the yearly positions of the peak of magnetic field $B$ (first row) and proton $n_{p}$ (second row) versus sunspot numbers (SSN). The first (second) column represents the results for the first (second) Gaussian curve. The linear fit equations are shown in the upper-left corner of each plot.}
    \label{fig:scatterPlot}
\end{figure}

The Pearson’s correlation coefficient, $r$, between $p_{B}^{1}$ or $p_{B}^{2}$ and the SSN are above 0.7, and between $p_{N}^{1}$ or $p_{N}^{2}$ and the SSN are above 0.65. Therefore, the magnetic field strength and the proton density of both slow and fast solar wind contributions are strongly related to the solar cycle.
Figure \ref{fig:scatterPlot} shows the scatter plots of the yearly position of the peak for $B$ and $n_{p}$ versus the sunspot number along with the linear fit.
For $p_{V}^{1}$, the correlation coefficient is 0.5, indicating a weak correlation. No correlation appears in other cases, as the correlation coefficients are below 0.1.

\section{Discussion}

In the previous section the Bi-Gaussian function was applied to reproduce the bulk solar wind at 1 AU (see Fig. \ref{fig:biGaussian_all}). The main solar wind parameters measured by ACE are properly reproduced by the Bi-Gaussian function, showing that the bulk solar wind can be described using a bi-modal approach. This result disagrees with the description of the solar wind as a simple statistical structure originating from the dynamical evolution and interaction of the flows at 1 AU \cite{Burlaga1999,Burlaga2001}.

The values for the parameters of the slow and fast components of the solar wind from the Bi-Gaussian fit (Table \ref{tab:fit_values_results}) agree with previous values for the slow and fast contributions  to solar wind (see Table \ref{tab:solar_wind_contributions}) provided by \cite{Bothmer2007}. Nevertheless, in our results, the fast contribution to solar wind is slightly colder. When comparing with the values from \cite{Hansteen2010}, we obtain a slower and colder  fast wind. Regarding magnetic field strength, our results in Table \ref{tab:weighted values} show two different values for the slow and fast contributions to solar wind, as in \cite{Bothmer2007}, while there is no difference in this parameter for the two contributions in the results by \cite{Hansteen2010}. 

We checked the bi-modal distribution of the solar wind at 1 AU, not only by using the ACE data set from its entire mission as a whole, but also the yearly data sets. The weighted average of the position of the peaks of the first and the second yearly Gaussian curves ($p_{1}, p_{2}$) of the different solar wind parameters (Table \ref{tab:weighted values}) agree with the results from the whole mission.

Considering that the average values have an uncertainty of $\pm \sigma$ (shadowed areas in Fig. \ref{fig:solarCycle_all}), we notice that several data points appear out of the shadowed areas. These outliers need to be carefully analysed. Nevertheless, no relationship is perceived between these outliers and the availability of data from ACE during the corresponding year.

In the case of magnetic field strength and proton density, $p_{1}$ and $p_{2}$ show strong dependence on the SSN. Also, the departure from the shadowed areas for $B$ and $n_{p}$ can be explained considering their relationship with solar cycle. Therefore, instead of considering a value for the magnetic field and the proton density with a larger uncertainty (e.g. $\pm 2 \sigma$), would be useful to obtain a value for slow and fast contributions, not for any time, but for different stages of the solar cycle. 

By recalculating for $B$ and $n_{p}$, the weighted averages for the two types of wind and different phases of solar cycle (maximum and minimum), we obtain the results that appear in Table \ref{tab:final_results} (columns 3 and 4). The years included in the average for the solar maximum comprise 1998 to 2003 and 2011 to 2016. Years from 2006 to 2010 are included in the period to compute the value during the solar minimum. The clear difference obtained in the typical value for $B$ in the different phases of the solar cycle reinforces the results obtained for the whole data sample showing a different typical value for $B$ for slow and fast contributions of solar wind, in agreement with \cite{Bothmer2007}  

In the case of those solar wind parameters where no relationship with the solar appears, that is $V_{p}$ and $T_{p}$, even if we extend the shadowed areas to $\pm 2 \sigma$, some points still remain outside the shadowed areas. The years corresponding to these outliers are 2009 and 2003. Therefore, we double checked the bi-Gaussian fittings for these years. 

\begin{figure}
	\includegraphics[width=\columnwidth]{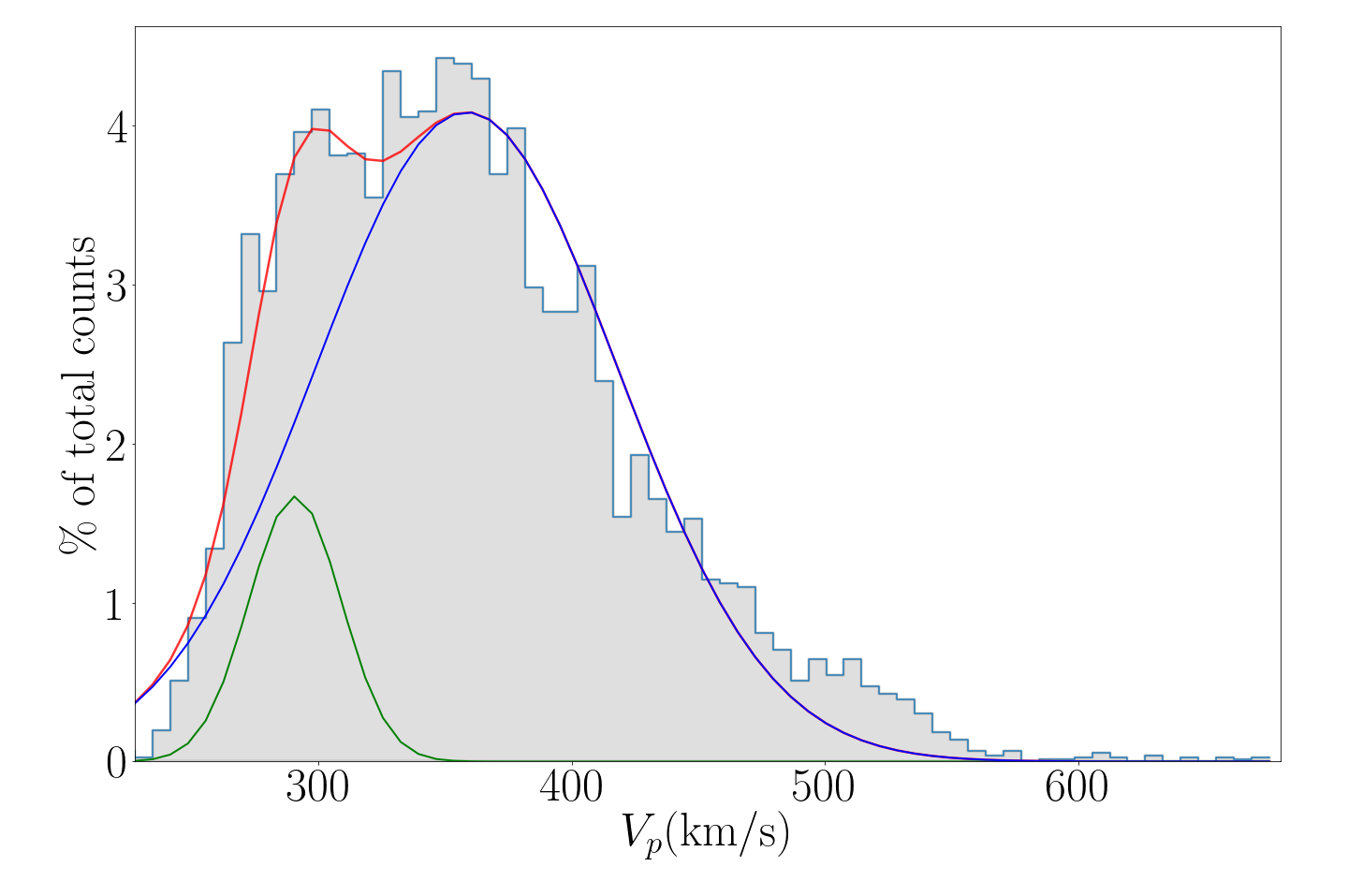}
    \caption{Empirical distribution functions of the solar wind speed, $V_{p}$, recorded by ACE during year 2009 and the fitting to a bi-Gaussian (red). Green and blue lines correspond to the single Gaussian curves.}
    \label{fig:V2009}
\end{figure}

Figure \ref{fig:V2009} shows the bi-Gaussian fitting for solar wind speed for the year 2009. Contrary to expectations, the height of the peak of the first Gaussian curve is very small when compared to that of the second one. During this year, the Sun was extremely quiet and the amount of flux from coronal holes was negligible. Therefore, the Gaussian with the greatest height, namely the second one, is the one corresponding to the slow solar wind for 2009. Moreover, the first Gaussian can be considered in this case as a mathematical artifact. Indeed, the fitting to a bi-Gaussian of the PDF for the speed in 2009 does not significantly improve the fit to a unique Gaussian. 

\begin{figure}
	\includegraphics[width=\columnwidth]{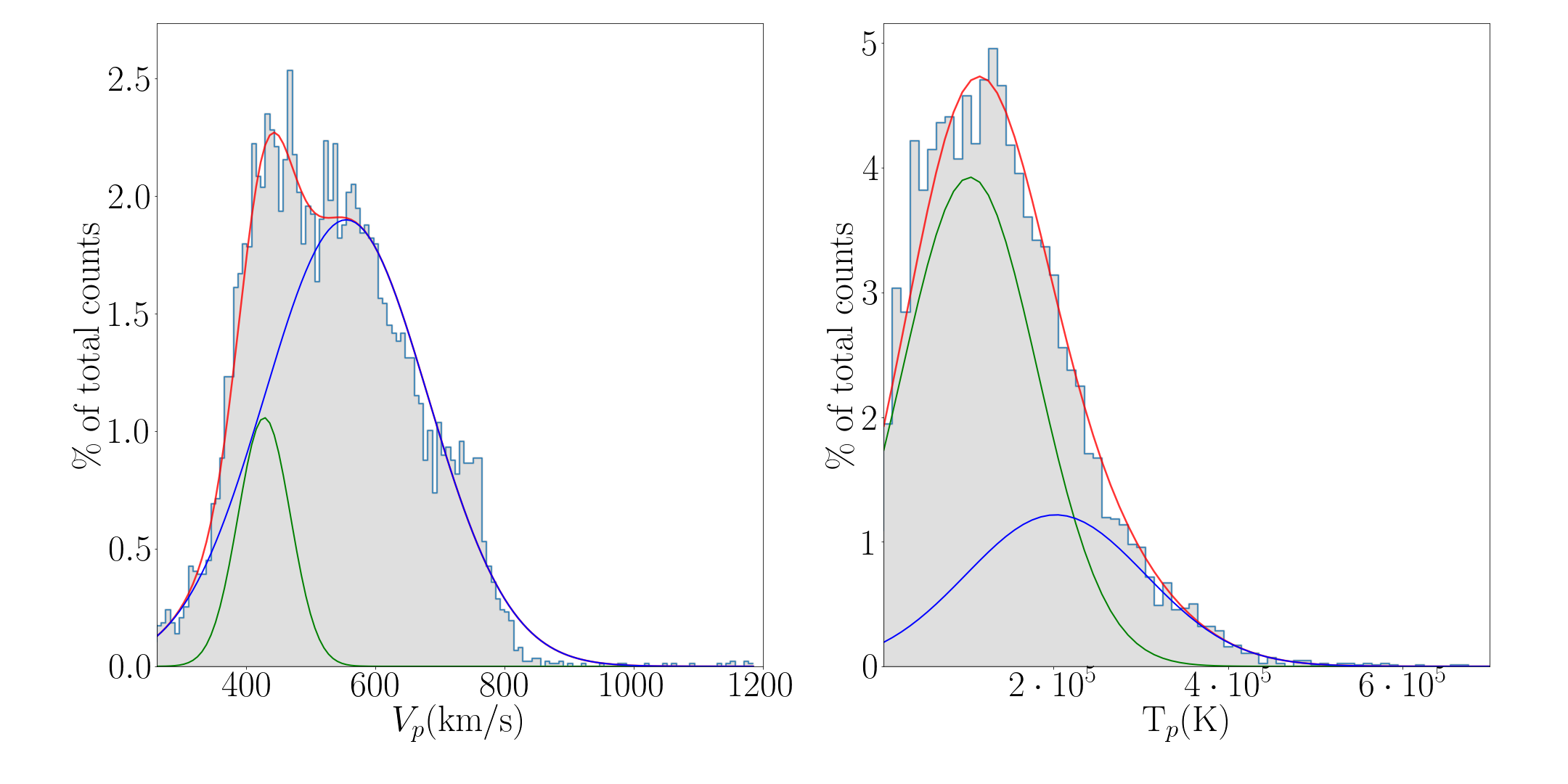}
    \caption{Empirical distribution functions of the solar wind speed, $V_{p}$, (left) and proton temperature $T_{p}$ (right) recorded by ACE during year 2003 and the fitting to a bi-Gaussian (red). Green and blue lines correspond to the single Gaussian curves.}
    \label{fig:T2003}
\end{figure}

A different example is that of the year 2003. During that year the Sun was extremely active with a major contribution from fast streams and from solar ejections. Moreover, interaction between different transients happened very often during this year. As a result, other types of wind different from the slow and the fast one contribute significantly to the PDF. The effect of this new population results in an increase of the value of the centre of the peak of both Gaussian curves for proton temperature and solar wind speed (Fig. \ref{fig:T2003}). Thus, this increase in $p_{1}$ and $p_{2}$ should not be associated to a different value for the typical $V_{p}$ or $T_{p}$ due to the slow and fast wind but to a spurious contribution.

After the deep analysis of the Bi-Gaussian curves obtained for the years 2003 and 2009 for $V_{p}$ and $T_{p}$, we conclude that an uncertainty of $\pm 2\sigma$ for the weighted average of $p_{1}$ and $p_{2}$ will be appropriate to describe typical values for these solar wind parameters. Therefore, we recap the typical values for solar wind in Table \ref{tab:final_results}

\section{Conclusions}
Here we show that the Bi-Gaussian function reproduces the bulk solar wind at 1 AU, not only for proton speed, but also for density, temperature and the magnetic field magnitude. This result suggests that the bulk solar wind at 1 AU is bi-modal, with a fast and a slow component. The typical values for the parameters of the two components are summarised in Table \ref{tab:final_results}

\begin{table}[h]
    \caption{Typical values for the main parameters of the slow and fast component of the solar wind according to the results from this study. The years included as the solar maximum (minimum) period are from 1998 to 2003 and from 2011 to 2016 (from 2006 to 2010).}
    \label{tab:final_results}
    \centering
    \begin{tabular}{lcccc} 
    \hline\hline
     & $V_{p} (km/s)$ & $B (nT)$ & $n_{p} (cm^{-3})$  & $T_{p} (\times 10^{5} K)$  \\      \hline
    Slow wind   & $380\pm40$    & $4.8\pm0.7$\footnote{M}   & $7\pm1^{2}$  & $0.5\pm0.6$\\   
    & & $3.3\pm0.3$\footnote{m}   & $4.8\pm0.5^{3}$ & \\
    \hline
    Fast wind   & $500\pm100$   & $7\pm1^{2}$   & $3.5\pm0.8^{2}$ & $1.0\pm0.8$\\    &  & $5.6\pm0.6^{3}$    & $2.4\pm0.2^{3}$ & \\  
    \hline
    \end{tabular}
    \end{table}

A bi-modal solar wind at 1 AU can be explained as emerging from a bi-modal source. This result leads us to the following open question: is there a clear boundary between small coronal holes responsible for slow solar wind and large coronal holes responsible for fast wind?

In the near future, new data from Parker Solar Probe will provide more in-situ observations of the solar wind in the inner heliosphere at different distances of up to 10 solar radii, shortly after the wind leaves the Sun. These new measurements will allow us to analyse the dynamical evolution of the different parameters of the bi-modal solar wind and to answer the above open question.

\section{Acknowledgements}
    This work was supported by the MINECO project AYA2016-80881-P (including FEDER funds). We thank the ACE SWEPAM and MAG instruments teams and the ACE Science Center for providing the ACE data. We also acknowledge WDC-SILSO, Royal Observatory of Belgium, Brussels for providing the Sunspot Number. C.L. acknowledges to Rahul Sharma for his help with the layout of the figures.

\bibliographystyle{apalike}  
\bibliography{references}

\end{document}